\documentclass{aa}  
\usepackage{graphicx}
\usepackage{txfonts}

\usepackage{natbib}
\bibpunct{(}{)}{;}{a}{}{,} 

\begin{document}
   \title{Parallaxes and physical properties of 11 mid-to-late T dwarfs}

   \author{F. Marocco\inst{1,2}
          \and
          R.L. Smart\inst{1}
          \and
          H.R.A. Jones\inst{3}
          \and
          B. Burningham\inst{3}
          \and
          M.G. Lattanzi\inst{1}
          \and
          S.K. Leggett\inst{4}
          \and
          P.W. Lucas\inst{3}
          \and
          C.G. Tinney\inst{5}
          \and
          A. Adamson\inst{6}
          \and
          D.W. Evans\inst{7}
          \and
          N. Lodieu\inst{8,9}
          \and
          D.N. Murray\inst{3}
          \and
          D.J. Pinfield\inst{3}
          \and
          M. Tamura\inst{10}
}

   \offprints{smart@oato.inaf.it}

   \institute{
   INAF/Osservatorio Astronomico di Torino,
              Strada Osservatorio 20, 10025 Pino Torinese, Italy
         \and
   Universit\`a degli Studi di Torino, Facolt\`a di Scienze MFN - Dipartimento di Fisica, 
              Via Pietro Giuria 1, 10126 Torino, Italy      
         \and
   Centre for Astrophysics Research, Science and Technology Research Institute,
             University of Hertfordshire, Hatfield AL10 9AB
          \and
   Gemini Observatory, Northern Operations Center, 670 N. A'ohoku Place, Hilo, HI 96720
          \and
   Anglo-Australian Observatory, P.O. Box 296, Epping, NSW1710, Australia       
          \and
   Joint Astronomy Centre, 660 North A'ohoku Place, Hilo, HI 96720, USA
          \and
   Institute of Astronomy, Madingley Road, Cambridge CB3 0HA, UK
          \and
   Instituto de Astrof\'isica de Canarias, V\'ia L\'actea s/n, E-38205 La Laguna, Tenerife, Spain
          \and
   Departamento de Astrof\'isica, Universidad de La Laguna (ULL), E-38205 La Laguna, Tenerife, Spain
          \and
   National Astronomical Observatory of Japan, 2-21-1 Osawa, Mitaka, Tokyo 181-8588, Japan
             }

   \date{Received 15/07/2010, accepted 20/08/2010}
 
    \abstract
    {}
    {We present parallaxes of 11 mid-to-late T dwarfs observed
      in the UKIRT Infrared Deep Sky Survey. We use these results to test the
      reliability of model predictions in magnitude-color space, determine
      a magnitude-spectral type calibration, and, estimate a bolometric
      luminosity and effective temperature range for the targets.}
    {We used observations from the UKIRT WFCAM instrument pipeline processed 
      at the Cambridge Astronomical Survey Unit. The
      parallaxes and proper motions of the sample were calculated using
      standard procedures. The bolometric luminosity was
      estimated using near- and mid-infrared observations with two different
      methods. The corresponding effective temperature ranges were found
      adopting a large age-radius range.}
    {We show the models are unable to predict the colors of the latest T
      dwarfs indicating the incompleteness of model opacities for NH$_3$, 
      CH$_4$ and H$_2$ as the temperature declines. We report the effective
      temperature ranges obtained.}
    {}


   \titlerunning{Properties of 11 T dwarfs}

   \maketitle

%

\section{Introduction}

Since the first discovery of a T dwarf by \citet{1995Natur.378..463N}
understanding the atmospheric processes and the role of chemical composition
for such low temperature objects has been very challenging. This remains an
important goal, particularly given that we consider these objects to be the link
between stars and giant exoplanets, therefore, their properties offer insights
into formation and evolution of planetary systems.

One of the fundamental parameters required to determine the physical
properties and to constrain theoretical models of celestial objects is
distance. Until now, only a few late T dwarfs have known distances: Wolf940B
\citep{1980AJ.....85..454H}, HD3651B \citep{1997A&A...323L..49P}, 2MASS
J0415-0935 \citep{2004AJ....127.2948V}, 2MASS J0939-2448
\citep{2008ApJ...689L..53B}, ULAS~J003402.77-005206.7
\citep{2010A&A...511A..30S}. Thus the theoretical models have been constrained
by the earlier T dwarfs. The UKIRT Infrared Deep Sky Survey \citep[hereafter
  UKIDSS]{2007MNRAS.379.1599L} is revealing large numbers of new T dwarfs
\citep{2010MNRAS.tmp..821B,2009MNRAS.397..258L, 2009MNRAS.395.1631L,
  2008MNRAS.391..320B, 2008MNRAS.390..304P, 2008MNRAS.385L..53C,
  2007MNRAS.379.1423L, 2007A&A...466.1059K} and in particular cool T8/T9
dwarfs.  An important follow up for the coolest T dwarfs observed in the
UKIDSS is the determination of their parallax. As shown in \citet[hereafter
  SJL10]{2010A&A...511A..30S} for ULAS~J003402.77-005206.7 using the UKIDSS
discovery image it is possible to reduce the time required to determine a
preliminary parallax and here we present 10 additional objects.

In this contribution all targets will be referred to by the discovery acronym
and right ascension short format hence ULAS~J003402.77-005206.7 becomes ULAS
0034, the full-names are given in Table \ref{photometry}. In Section 2 we
describe the observations and procedures and in Section 3 we report the
results for the 11 targets. In Section 4 we compare these results to current
models and in Section 5 we calculate L$_{bol}$ and T$_{eff}$ of the objects in
the sample. Finally in Section 6 we discuss the results obtained.

\section{Observations and Reduction Procedures }

The observations for the parallax determination began in 2007 and the target
list of 11 objects was drawn from confirmed T dwarfs observed at that time 
in the UKIDSS Large Area Survey.
In Table \ref{photometry} we list the objects along with MKO magnitudes
\citep{2002PASP..114..180T} and near-infrared (hereafter NIR) spectral types.
The NIR spectral type classification follows the scheme described in
\citet{2006ApJ...637.1067B}. The only object with an already published
parallax was SDSS~0207 which was included because the USNO parallax had a
large relative error \citep[$\sim$28\%,][]{2004AJ....127.2948V}.

The procedures for observing, image treatment and parallax determination
follow those described in SJL10. In that paper the parallax observations of
ULAS~0034 maintained the same target position as the discovery image and were
aligned using a simple linear transformation.  Assuming the astrometric
distortion pattern does not change, the use of the discovery image as part of
the parallax solution allows a much shorter dedicated observational campaign
to obtain a precise parallax. For CFBDS~0059, SDSS~0207, ULAS~0948
and ULAS~2239 this was not possible because the target was very close to a 
chip edge in the first UKIDSS image. The WFCAM astrometric distortion is significant, so 
moving the reference frame on the focal plane results in poorer astrometric
transformations (i.e. larger residuals in the solutions) and so in this 
situation those frames that are significantly offset are given lower weight. 

Initially all observations were made with the same total exposure time of 400
seconds. For this exposure, in the case of the anonymous stars in the field of 
ULAS~0034, the centroiding precision is a constant 20mas until around \textit{J}=18.4 and 
deteriorates quickly to 60mas at \textit{J}=19.2 (Figure 2 in SJL10). To 
ensure that centroiding precision is optimal even in poor observing conditions 
we have increased the exposure times to 710s for the faint targets (ULAS~0034, 
0948, 1150, 1315, 2239 and CFBDS~0059). The precision of the final solutions 
for the faint targets reflects the poor quality of some of the earlier 400 
second observations.

\begin{figure}
  \centering
  \includegraphics[width=8cm]{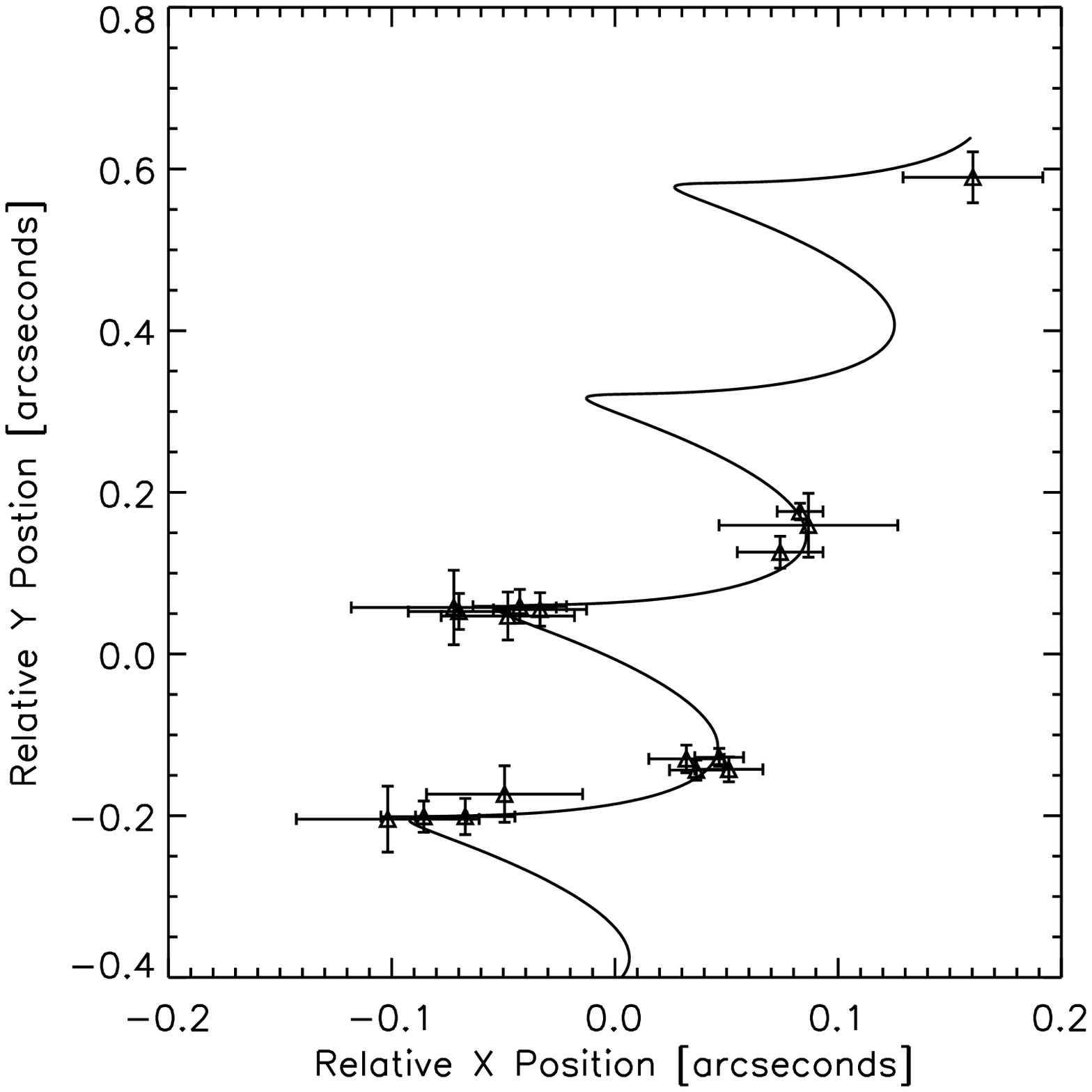}
  \includegraphics[width=8cm]{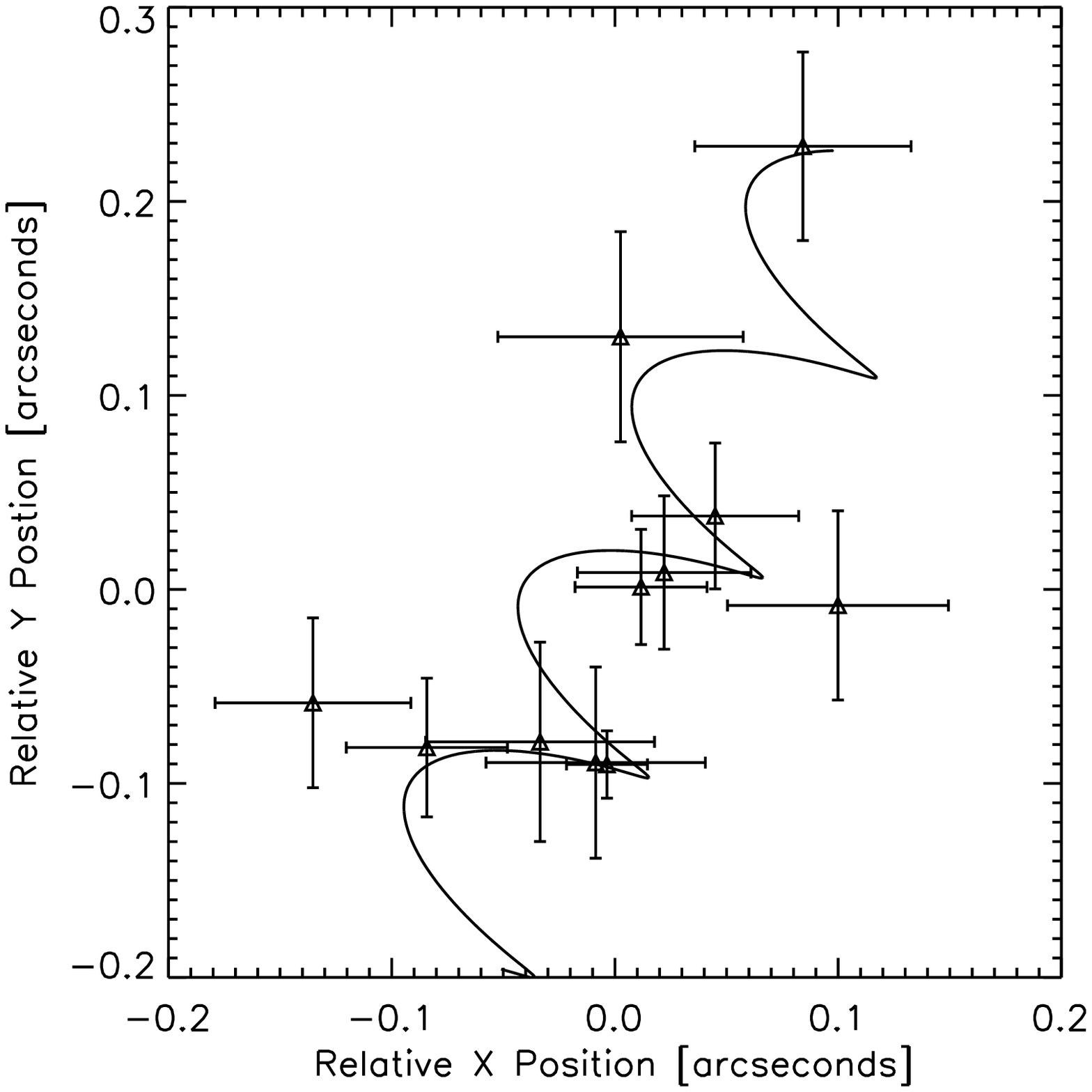}
  \caption{Observations for the targets ULAS~0901 (top panel) and 1315 (bottom panel). 
  The highest point is the discovery image. The observational history and distance of 
  these two targets are similar but they differ by a magnitude in apparent brightness. 
  The solution shows the effect of low signal-to-noise observations in the beginning 
  of the parallax sequence.}
  \label{twosolutions}
\end{figure}

In figure \ref{twosolutions} we plot the solution of ULAS~0901 (top panel) and 1315 (bottom panel) which
are objects of similar distances and observational history but with magnitudes
of 17.90 and 18.86 respectively, e.g. straddling the precision borderline, to
show high and low quality solutions.

\begin{table*}
\begin{center}
\caption{Infrared magnitudes and spectral types of the 11 targets.}
\begin{tabular}{l|c|c|c|c|c|c|c}
Full Name & \textit{z}$_{AB}$ & \textit{Y} & \textit{J} & \textit{H} & \textit{K} & Sp. Type & Refs (D,P,T) \\
\hline    	  
ULAS J003402.77-005206.7 & 22.11 $\pm$ 0.05 & 18.90 $\pm$ 0.10 & 18.15 $\pm$ 0.03 & 18.49 $\pm$ 0.04 & 18.48 $\pm$ 0.05 & T9   & 1,1,6 \\
CFBDS J005910.90-011401.3& 21.93 $\pm$ 0.05 & 18.82 $\pm$ 0.02 & 18.06 $\pm$ 0.03 & 18.27 $\pm$ 0.05 & 18.63 $\pm$ 0.05 & T9   & 2,2,6 \\
SDSS J020742.48+000056.2 & 20.11 $\pm$ 0.60 & 17.94 $\pm$ 0.03 & 16.75 $\pm$ 0.01 & 16.79 $\pm$ 0.04 & 16.71 $\pm$ 0.05 & T4.5 & 3,(4,7),8 \\
ULAS J082707.67-020408.2 & \ldots           & 18.29 $\pm$ 0.05 & 17.19 $\pm$ 0.02 & 17.44 $\pm$ 0.05 & 17.52 $\pm$ 0.11 & T5.5 & 4,4,4 \\
ULAS J090116.23-030635.0 & \ldots           & 18.82 $\pm$ 0.05 & 17.90 $\pm$ 0.04 & 18.46 $\pm$ 0.13 & $>$ 18.21        & T7.5 & 4,4,4 \\
ULAS J094806.06+064805.0 & \ldots           & 20.03 $\pm$ 0.14 & 18.85 $\pm$ 0.07 & 19.46 $\pm$ 0.22 & $>$ 18.62        & T7   & 4,4,4 \\
ULAS J101821.78+072547.1 & \ldots           & 18.90 $\pm$ 0.08 & 17.71 $\pm$ 0.04 & 17.87 $\pm$ 0.07 & 18.12 $\pm$ 0.17 & T5   & 4,4,4 \\
ULAS J115038.79+094942.8 & 22.44 $\pm$ 0.10 & 19.92 $\pm$ 0.08 & 18.68 $\pm$ 0.04 & 19.23 $\pm$ 0.06 & 19.06 $\pm$ 0.05 & T6.5p& 5,5,5 \\
ULAS J131508.42+082627.4 & 22.82 $\pm$ 0.10 & 20.00 $\pm$ 0.08 & 18.86 $\pm$ 0.04 & 19.50 $\pm$ 0.10 & 19.60 $\pm$ 0.12 & T7.5 & 5,5,5 \\
ULAS J133553.45+113005.2 & 22.04 $\pm$ 0.10 & 18.81 $\pm$ 0.04 & 17.90 $\pm$ 0.01 & 18.25 $\pm$ 0.01 & 18.28 $\pm$ 0.03 & T9   & 6,6,6 \\
ULAS J223955.76+003252.6 & \ldots           & 19.94 $\pm$ 0.17 & 18.85 $\pm$ 0.05 & 19.10 $\pm$ 0.10 & 18.88 $\pm$ 0.06 & T5.5 & 4,4,4 \\
\hline
\end{tabular}
\label{photometry}
\end{center}
\small{\textit{YJHK} are in the MKO Vega photometric system, while \textit{z} is in AB system. The uncertainty in the spectral type is $\pm$0.5.\\
       References (D = Discovery, P = Photometry, T = Spectral Type): 
         1- \citet{2007MNRAS.381.1400W}
         2- \citet{2008AA...482..961D}
         3- \citet{2002ApJ...564..466G}
         4- \citet{2007MNRAS.379.1423L}
         5- \citet{2008MNRAS.390..304P}
         6- \citet{2008MNRAS.391..320B}
         7- \citet{2004AJ....127.3553K}
         8- \citet{2006ApJ...637.1067B}}
\end{table*}

\section{Results}

The astrometric parameters derived for the 11 targets are in Table
\ref{results}. For each one we report: target name, position (J2000), number
of reference stars, number of observations used, absolute parallax, proper
motion components, tangential velocity, the time span covered by the
observations and the relative-to-absolute parallax correction applied.The
proper motion of the targets have all been brought to an absolute system using
the galaxies in the field.

\begin{table*}
\begin{center}
\caption{Parallaxes and proper motions of the 11 targets.}
\leavevmode
\begin{tabular}{l|c|c|c|c|c|c|c|c|c}
 Target    & $\alpha$ & $\delta$ & N*, obs & $\pi^{abs}$ $\pm$ $\sigma_{\pi}$ & $\mu_{\alpha}^{abs}$ $\pm$ $\sigma_{\mu_{\alpha}}$ & $\mu_{\delta}^{abs}$ $\pm$ $\sigma_{\mu_{\delta}}$ & V$_{tan}$ & Time span & COR \\
 	&	(h:m:s) & (d:m:s) & & (mas) & (mas/y) & (mas/y) & (km/s) & (years) & (mas) \\
\hline 
   ULAS~0034  &  0:34:02.7 & - 0:52:07.8 & 135, 17 &  78.0 $\pm$ 3.6 &  -18.5 $\pm$  3.2 & -363.3 $\pm$  3.6 & 21.7 $\pm$  1.0 & 3.81 & 1.24 \\
   CFBDS~0059 &  0:59:10.9 & - 1:14:01.4 &  70, 13 & 108.2 $\pm$ 5.0 &  878.8 $\pm$  8.4 &   50.5 $\pm$  4.8 & 38.6 $\pm$  1.8 & 2.69 & 1.19 \\
   SDSS~0207  &  2:07:42.9 & + 0:00:56.0 &  47, 17 &  29.3 $\pm$ 4.0 &  158.8 $\pm$  3.0 &  -14.3 $\pm$  3.9 & 25.8 $\pm$  3.6 & 3.70 & 1.37 \\
   ULAS~0827  &  8:27:07.6 & - 2:04:08.4 & 418, 17 &  26.0 $\pm$ 3.1 &   26.8 $\pm$  2.7 & -108.9 $\pm$  2.3 & 20.5 $\pm$  2.5 & 3.94 & 0.87 \\
   ULAS~0901  &  9:01:16.2 & - 3:06:35.4 & 241, 19 &  62.6 $\pm$ 2.6 &  -38.6 $\pm$  2.3 & -261.2 $\pm$  2.8 & 20.0 $\pm$  0.8 & 3.90 & 1.00 \\
   ULAS~0948  &  9:48:06.1 & + 6:48:04.5 & 152, 15 &  27.2 $\pm$ 4.2 &  199.4 $\pm$  7.0 & -273.9 $\pm$  6.2 & 59.1 $\pm$  9.3 & 1.58 & 0.98 \\
   ULAS~1018  & 10:18:21.7 & + 7:25:46.8 & 198, 14 &  25.0 $\pm$ 2.0 & -183.7 $\pm$  2.6 &  -15.1 $\pm$  3.1 & 34.9 $\pm$  2.8 & 2.00 & 1.04 \\
   ULAS~1150  & 11:50:38.7 & + 9:49:42.8 & 105, 10 &  16.8 $\pm$ 7.5 & -107.6 $\pm$ 17.1 &  -31.9 $\pm$  4.5 & 31.7 $\pm$ 14.9 & 2.91 & 1.09 \\
   ULAS~1315  & 13:15:08.4 & + 8:26:27.0 & 213, 11 &  42.8 $\pm$ 7.7 &  -60.2 $\pm$  8.3 &  -95.8 $\pm$ 10.0 & 12.5 $\pm$  2.5 & 3.07 & 0.95 \\
   ULAS~1335  & 13:35:53.4 & +11:30:05.1 & 196,  8 &  96.7 $\pm$ 3.2 & -196.9 $\pm$  4.9 & -201.0 $\pm$  6.3 & 13.8 $\pm$  0.5 & 2.18 & 0.95 \\
   ULAS~2239  & 22:39:55.7 & + 0:32:52.7 & 120, 15 &  10.4 $\pm$ 5.2 &  125.3 $\pm$  5.4 & -108.4 $\pm$  5.2 & 75.7 $\pm$ 37.8 & 2.95 & 0.96 \\
\hline
\end{tabular}
\label{results}
\end{center}
\small{In the fourth column we report the number of reference stars (N*) and
  the number of observations (obs). In the ninth column we report the time
  span covered by the observations and in the last one the
  relative-to-absolute parallax correction (COR).}
\end{table*} 

The relative errors for most of the targets are less than
10$\%$. Observations are continuing and we hope to be able to reduce them to
5$\%$ by the end of the campaign. Significant exceptions are ULAS~1150, 1315
and 2239 which are also amongst the faintest objects under study, hence
degraded the most by the borderline signal-to-noise in the early
observations. In addition, parallax observations of ULAS~2239 were shifted
from the discovery image position lowering the weight of the first point.

In Table \ref{distances} we compare the astrometric distances obtained here with the 
estimated ones given in the discovery paper of each object. The estimations were made 
with different techniques and the reader is referred to the discovery papers (see 
footnote to Table \ref{photometry}) for further details. The discovery distance range 
is usually much larger than our measured one and is overestimated for the late-Ts. 
This comparison underlines the need for measured parallaxes that are 
model independent.

\begin{table}
\caption{The 1-sigma distance range obtained here compared to those estimated in the discovery papers.}
\begin{center}
\begin{tabular}{c|c|c|c}
Name & Astrometric & Discovery & Discovery \\
 & distance (pc) & distance (pc) & reference \\
\hline
ULAS~0034 & 12.2-13.4  & 14-24 & 1 \\
CFBDS~0059 & 8.8-9.6   & 8-18 & 2 \\
ULAS~0827 & 33.8-43.0  & 24-39 & 3 \\
ULAS~0901 & 15.3-16.7  & 21-33 & 3 \\
ULAS~0948 & 31.1-42.5  & 38-60 & 3 \\
ULAS~1018 & 36.8-43.2  & 33-52 & 3 \\
ULAS~1150 & 32.9-86.1  & 42-60 & 4 \\
ULAS~1315 & 19.2-27.6  & 34-48 & 4 \\
ULAS~1335 & 10.0-10.6  & 8-12 & 5 \\
ULAS~2239 & 48.0-144.2 & 52-83 & 3 \\
\hline
\end{tabular}
\end{center}
\label{distances}
\small{References:
         1- \citet{2007MNRAS.381.1400W}
         2- \citet{2008AA...482..961D}
         3- \citet{2007MNRAS.379.1423L}
         4- \citet{2008MNRAS.390..304P}
         5- \citet{2008MNRAS.391..320B}
       }
\end{table}

\section{Model Fitting}

\begin{figure*}
  \centering
  \includegraphics[width=14cm]{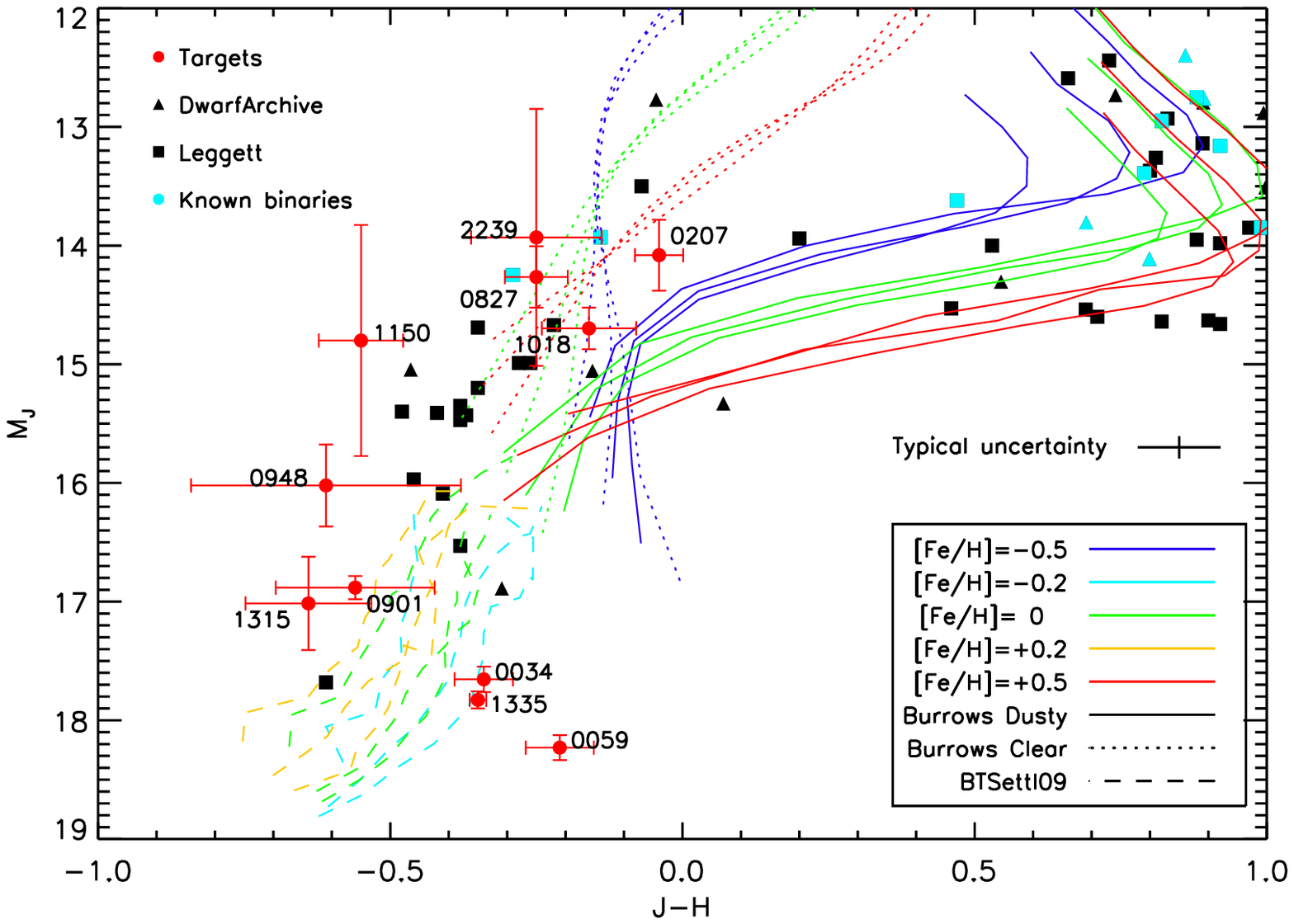}
  \includegraphics[width=14cm]{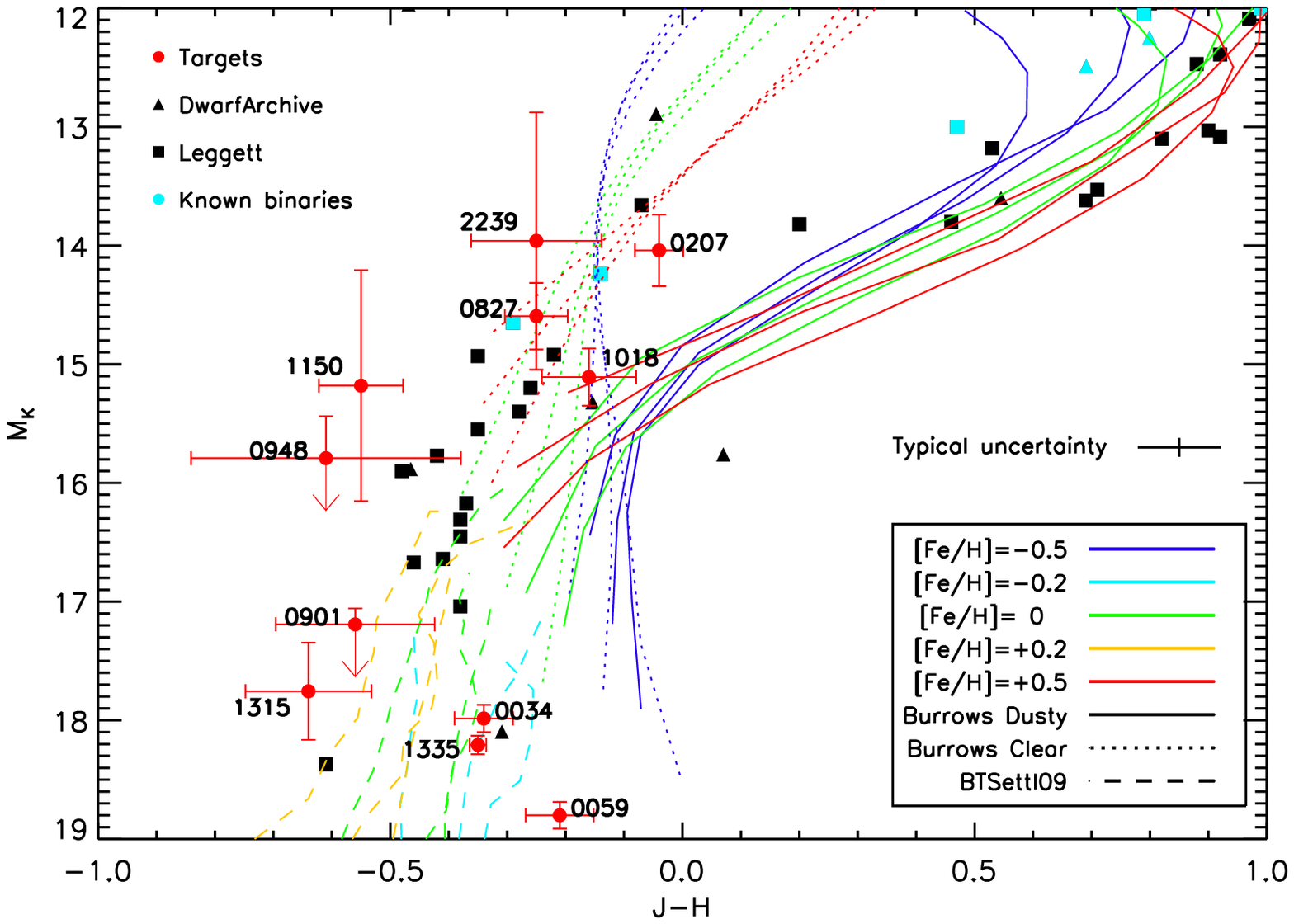}
  \caption{Color-magnitude diagrams for a sample of L and T dwarfs. The colored
    lines are theoretical tracks from BSH06 and BTSettl09, for different gravities 
    and metallicities. For each metallicity, the gravity increases from left to 
    right, assuming the values 4.5,5.0,5.5 for BSH06 tracks and 4.5,5.0,5.25 for 
    BTSettl09. The 11 targets presented here are plotted as filled circles with 
    their associated error bars, the Dwarf archive objects are plotted as triangles 
    and the Leggett archive objects as squares. All magnitudes in the MKO
    system. The cross above the legend indicates the typical
    uncertainty on the literature objects.}

  \label{mvsj-h+model}
\end{figure*}

\begin{figure*}
  \centering
  \includegraphics[width=14cm]{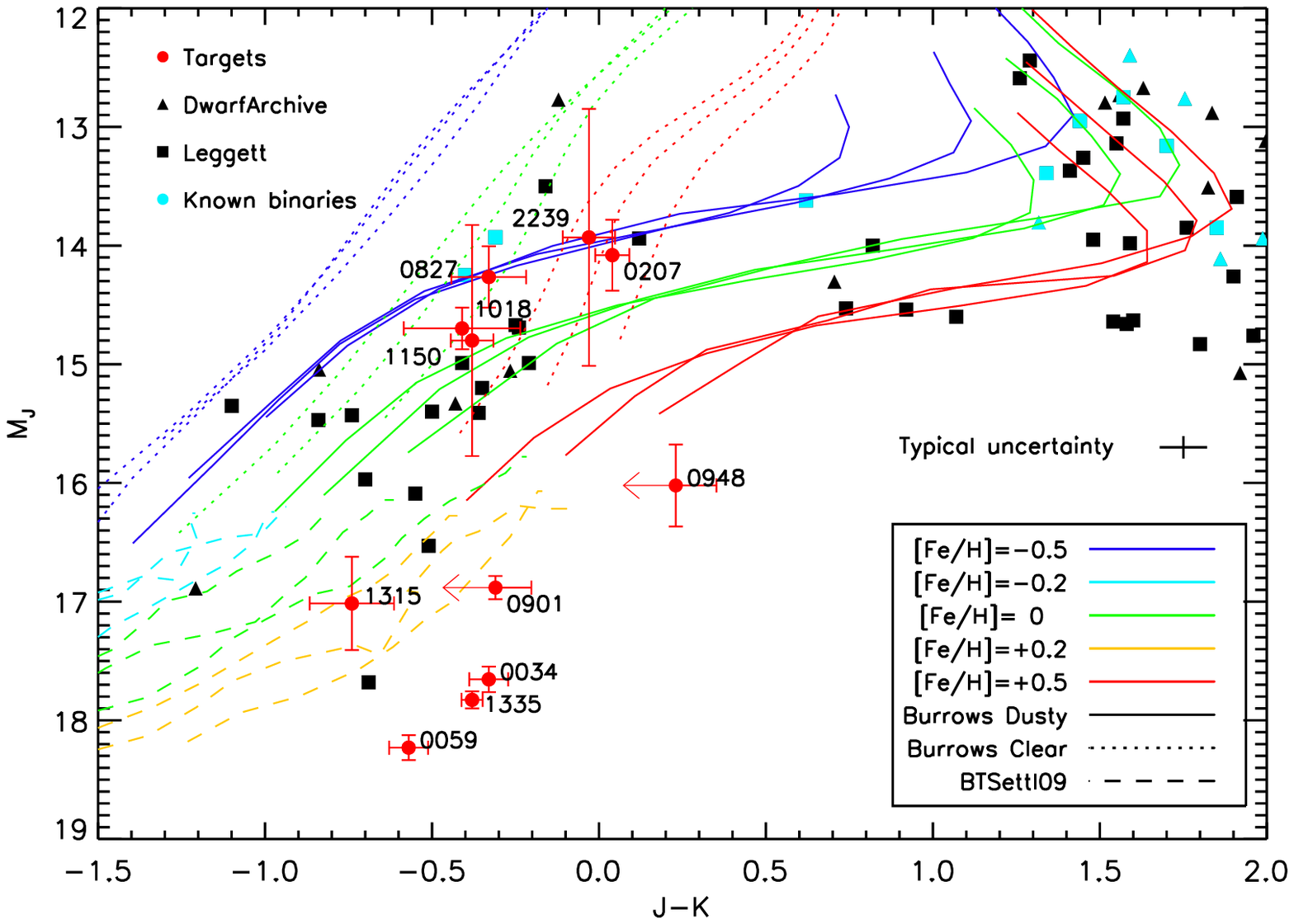}
  \includegraphics[width=14cm]{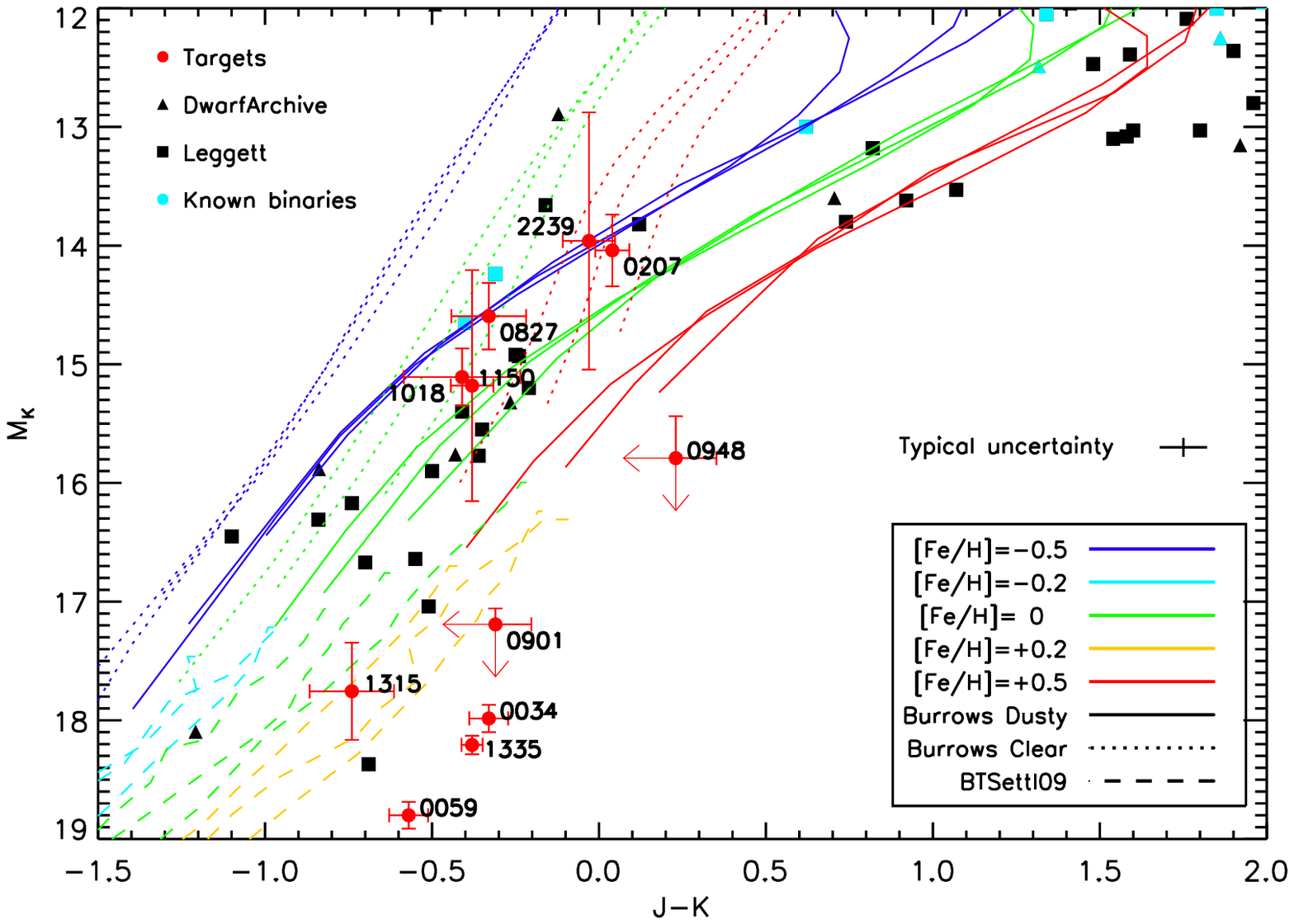}
  \caption{Same as Fig. \ref{mvsj-h+model} but with the color \textit{J-K} instead of
    \textit{J-H}. For each metallicity, the gravity increases from right to left, 
    assuming the values 4.5,5.0,5.5 for BSH06 tracks and 4.5,5.0,5.25 for 
    BTSettl09.}
  \label{mvsj-k+model}
\end{figure*}

In Figs. \ref{mvsj-h+model} and \ref{mvsj-k+model} we present four color -
absolute magnitude diagrams for a sample of L and T dwarfs with M$_J$ $>$ 12
and M$_K$ $>$ 12 respectively. The 11 targets are plotted as filled
circles. Parallaxes and magnitudes of the other objects are from two
on-line archives:\\
1) The L/T dwarf archive maintained by S. K. Leggett
(http://staff.gemini.edu/$\sim$sleggett/2010\_phot\_tab.txt, hereafter Leggett
archive): This archive contains a compendium of 225 objects with MKO \textit{YJHKL'M'} 
magnitudes and IRAC [3.55], [4.49], [5.73] and [7.87] magnitudes. Where used 
these objects are plotted as filled squares.\\
2) The M/L/T dwarf archive (www.dwarfarchives.org, hereafter Dwarf archive):
This is an on line compendium of all published L/T dwarfs and selected M
dwarfs. As of 10/01/2010 there were 752 L \& T dwarfs, reporting \textit{JHK}
magnitudes, parallaxes and proper motions. Where used these objects are
plotted as filled triangles.

To make Figs. \ref{mvsj-h+model} and \ref{mvsj-k+model} clearer 
we omitted the error bars on the literature objects. An indication of the 
typical uncertainty on these points is given by the black cross above 
the legend.

All magnitudes are in the MKO system and preferentially taken from the Leggett
archive as they are measured directly in that system. The majority of the
infrared magnitudes in the Dwarf archive are in the 2MASS system and when
needed we use the relations in \citet{2004PASP..116....9S} to convert to the MKO system.

Colored lines in Figs. \ref{mvsj-h+model} and \ref{mvsj-k+model} are the model
predictions by \citet[hereafter BSH06]{2006ApJ...640.1063B} and
\citet[hereafter BTSettl09]{2003A&A...411L.473A, 2007A&A...474L..21A,
2009A&A...506..993A}. BSH06 tracks covers the temperature range 700-2000 K, with 
log[g]=4.5,5.0,5.5 and [Fe/H]=-0.5,0,+0.5. BTSettl09 covers the range 
500-780 K, with log[g]=4.5,5.0,5.25 and [Fe/H]=-0.2,0,+0.2. For a given 
metallicity (indicated in the plots by a given color) log[g] increases from left 
to right in J-H, while it increases from right to left in J-K. 

Model colors and magnitudes were obtained by convolving the theoretical spectra with 
the UKIDSS filter profiles \citep{2006MNRAS.367..454H} to calculate fluxes. We 
interpolate the model spectra with a spline to have the same binning as the 
filter profiles, apply the selected profile, and integrate to obtain the total 
flux. The integrated fluxes were then converted into an absolute magnitude using 
as a zero point a Vega spectrum treated the same way.

The absolute magnitudes plotted in Figs. \ref{mvsj-h+model} and
\ref{mvsj-k+model} require from the models the flux at 10 pc. The BSH06 models
supply the flux at the surface of the object and at 10 pc, the latter
calculated assuming the radius-log(g)-T$_{eff}$ relation from
\citet{1997ApJ...491..856B}.  The BTSettl09 models provide only the flux at
the surface of the object, to find the flux at 10pc we assume the radii from
\citet{2003A&A...402..701B} corresponding to the model log(g) and effective
temperature.

In Figs. \ref{mvsj-h+model} and \ref{mvsj-k+model} we note
that the BTSettl09 tracks for high and low metallicity are swapped in the
different color spaces. In J-H space tracks for high metallicity are bluer
than the low metallicity ones, while in J-K they are redder. This follows the
trend suggested by BSH06 tracks for higher temperature, and can be seen also
in the models of \citet{2008ApJ...689.1327S}. Due to this swap, the predictions 
for the T9 dwarfs are incompatible in the two color spaces, i.e. they predict high
gravity - low metallicity in J-H and low gravity - high metallicity in
J-K. Moreover, the T9s are redder than predicted by the theoretical
models, especially in J-K. This failure may be due to an incorrect prediction
of the flux emitted in the H and K-band resulting from  known opacity
deficiencies in modeling the collision-induced
absorption of H$_2$ and the wing of the K I doublet resonance, and the lack 
of an appropriate list of absorption lines of CH$_4$ and NH$_3$ at such low 
temperatures \citep{2010ApJ...710.1627L}.

The predictions for earlier objects are consistent in the two color spaces, but 
in Fig. 2 they tend to slightly underestimate the magnitudes of the T6.5-T7s. 
Further discussion on the models predictions and individual objects is in
Section 6.

\begin{figure*}
  \centering
  \includegraphics[width=13.5cm]{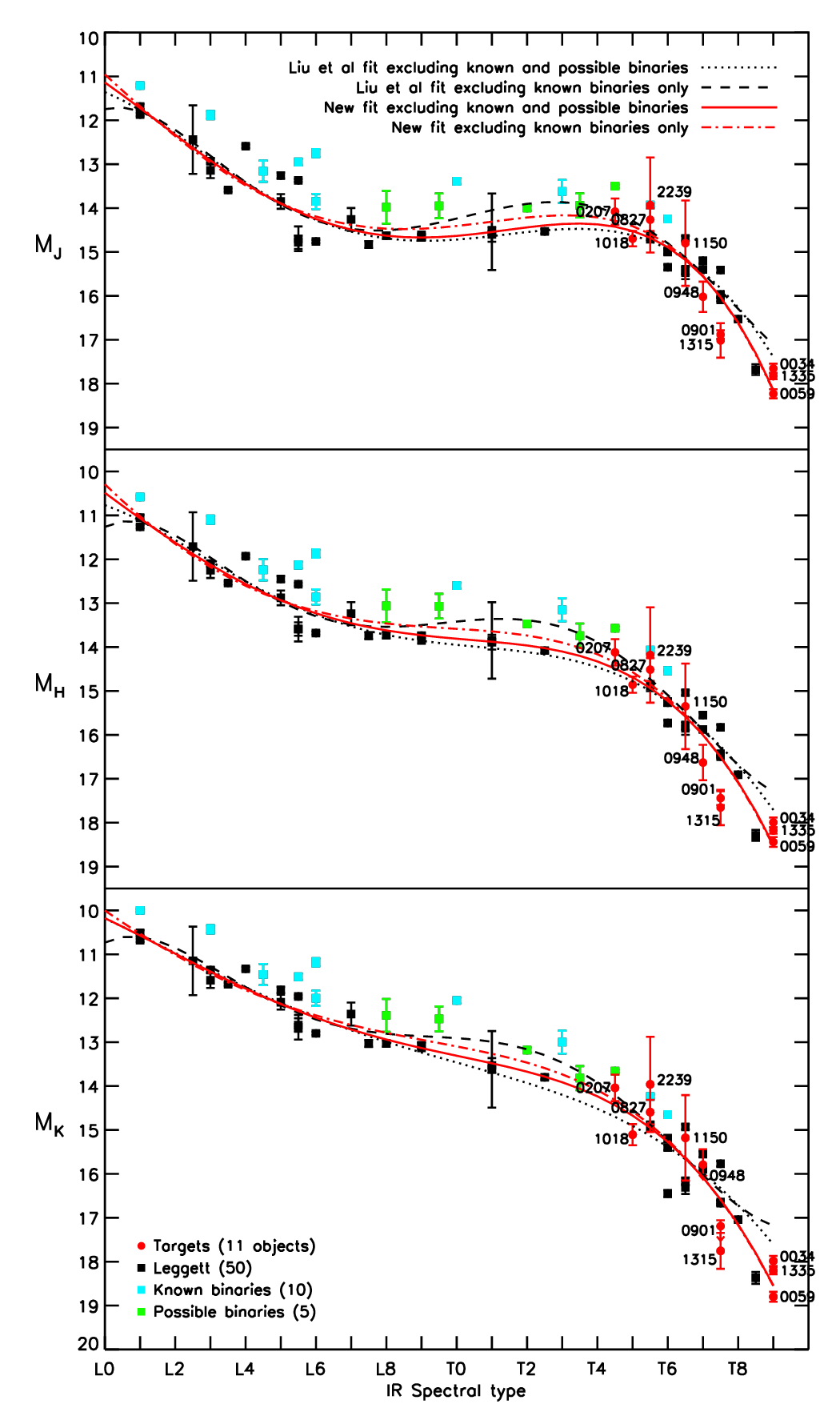}
  \caption{\textit{J}, \textit{H} and \textit{K}-band absolute magnitude as a function of IR spectral
    type. Labeled objects are those with new parallaxes presented here. Blue 
    squared points are known to be unresolved binaries, while green squared 
    points are possible unresolved binaries. The over plotted lines are polynomial 
    fit by \citet{2006ApJ...647.1393L} based on data tabulated by
    \citet{2004AJ....127.3553K}. The dotted line was obtained excluding all
    the known and possible binaries, the dashed one excluding only the known
    binaries. The red lines are our new polynomial fits.}
  \label{magvsirsptype}
\end{figure*}

In Fig. \ref{magvsirsptype} we present three absolute magnitude - IR spectral
type diagrams. The IR spectral type classification follows the scheme
described in \citet{2002ApJ...564..466G} for L dwarfs and the scheme described
in \citet{2006ApJ...637.1067B} for T dwarfs. The over plotted curves are
polynomial fits  derived in \citet{2006ApJ...647.1393L}, with the dotted
line representing the polynomial obtained excluding from the fit all the known
and possible binaries, while the dashed line is the one obtained excluding
only the known binaries. Possible binaries were selected by Liu et al. 
based on their relatively high T$_{eff}$ compared
to objects of similar spectral type in the \citet{2004AJ....127.3516G}
measurements. One of the possible binaries indicated by Liu et al. (SDSS J1021-0304) was
confirmed as a binary by \citet{2006ApJ...637.1067B}, so now is plotted as a
known binary. Following the convention used before, filled circles are the 11
targets and squares are objects from the Leggett archive, while blue objects are 
known binaries and green objects are possible binaries. The trend for T8.5 and 
T9 is an extrapolation of the Liu et al polynomial, since their sample consisted of 
objects between L0 and T8.

The 11 targets indicate a steeper trend in the sequence beyond T8 than the
extrapolation of the Liu et al. polynomial. In Fig. \ref{magvsirsptype} the
red lines are our 4-th order polynomials fit to the data including the new
objects presented here, but excluding the 14 objects without magnitude
naturally in the MKO system. This choice was made as the
\citet{2004PASP..116....9S} transformations from the 2MASS to the MKO system
were derived using hotter objects and employing them may introduce systematic
errors in the resulting fit of the cooler objects. In Table \ref{fitcoeff} we
list the coefficients and errors of the fit. The difference between the two
fits with and without possible binaries is reduced compared to Liu et al.,
probably because of the smaller statistical weight of the possible binaries in
this study (5 over a total sample of 61 objects) with respect to Liu et al. (6
over 43). As the sample of late T dwarfs is still small,
identification of possible binaries in this region may be incomplete.

We also tested polynomials from 3rd to 10th order but after 4th order there
was no significant improvement in the sigma of the fit.

\begin{table*}
\begin{center}
\caption{Coefficients of the polynomial fit.}
\begin{tabular}{c c c c c c}
    Magnitude & a$_0$ & a$_1$ & a$_2$ & a$_3$ & a$_4$ \\
\hline
\multicolumn{6}{c}{Excluding known and possible binaries} \\
\hline
   M$_J$ & 11.140 $\pm$ 0.052 & 0.556 $\pm$ 0.032 & 3.74 $\pm$ 0.62 $\times$10$^{-2}$ & -9.49 $\pm$ 0.45 $\times$10$^{-3}$ & 3.69 $\pm$ 0.11 $\times$10$^{-4}$ \\
   M$_H$ & 10.486 $\pm$ 0.051 & 0.590 $\pm$ 0.031 & -3.80 $\pm$ 0.61 $\times$10$^{-3}$ & -4.35 $\pm$ 0.44 $\times$10$^{-3}$ & 2.15 $\pm$  0.11 $\times$10$^{-4}$ \\
   M$_K$ & 10.180 $\pm$ 0.056 & 0.375 $\pm$ 0.035 & 2.31 $\pm$ 0.67 $\times$10$^{-2}$ & -5.03 $\pm$ 0.49 $\times$10$^{-3}$ & 2.11 $\pm$ 0.12 $\times$10$^{-4}$ \\
\hline   
\multicolumn{6}{c}{Excluding known binaries} \\
\hline   
   M$_J$ & 10.957 $\pm$ 0.053 & 0.736 $\pm$ 0.032 & -5.53 $\pm$ 0.61 $\times$10$^{-3}$ & -6.31 $\pm$ 0.44 $\times$10$^{-3}$ & 2.96 $\pm$ 0.11 $\times$10$^{-4}$ \\
   M$_H$ & 10.292 $\pm$ 0.052 & 0.778 $\pm$ 0.031 & -4.77 $\pm$ 0.60 $\times$10$^{-2}$ & -1.13 $\pm$ 0.43 $\times$10$^{-3}$ & 1.41 $\pm$  0.10 $\times$10$^{-4}$ \\
   M$_K$ & 10.005 $\pm$ 0.056 & 0.548 $\pm$ 0.034 & -1.84 $\pm$ 0.64 $\times$10$^{-2}$ & -1.93 $\pm$ 0.46 $\times$10$^{-3}$ & 1.38 $\pm$ 0.11 $\times$10$^{-4}$ \\
\hline
\end{tabular}
 \\
\end{center} 
\small{The fits are defined as in \citet{2006ApJ...647.1393L}: M$_{XX}$ =
  $\sum_{i=0}^{4}${ }a$_{i}${ }$\times${ }(SpT)$^{i}$ where XX indicates \textit{J}, 
  \textit{H} or \textit{K}-band magnitude and the spectral types are defined following 
  the   convention SpT=1 for L1, SpT=9 for L9, SpT=10 for T0 etc. The fit is valid for 
  spectral types from L0 to T9. Infrared spectral types are used for both L and T dwarfs.}
\label{fitcoeff}
\end{table*}

\section{Luminosity and Effective Temperature.}

Next we calculate an effective temperature range for our targets. To do this we used the classical Stefan-Boltzmann law:
\begin{equation}
L_{bol} = 4 \pi \sigma R^2 T_{eff}^4
\label{sb}
\end{equation}
hence to calculate the temperature we need to know the radius and the bolometric luminosity of each object.

\begin{figure}
  \centering
  \includegraphics[width=8.5cm]{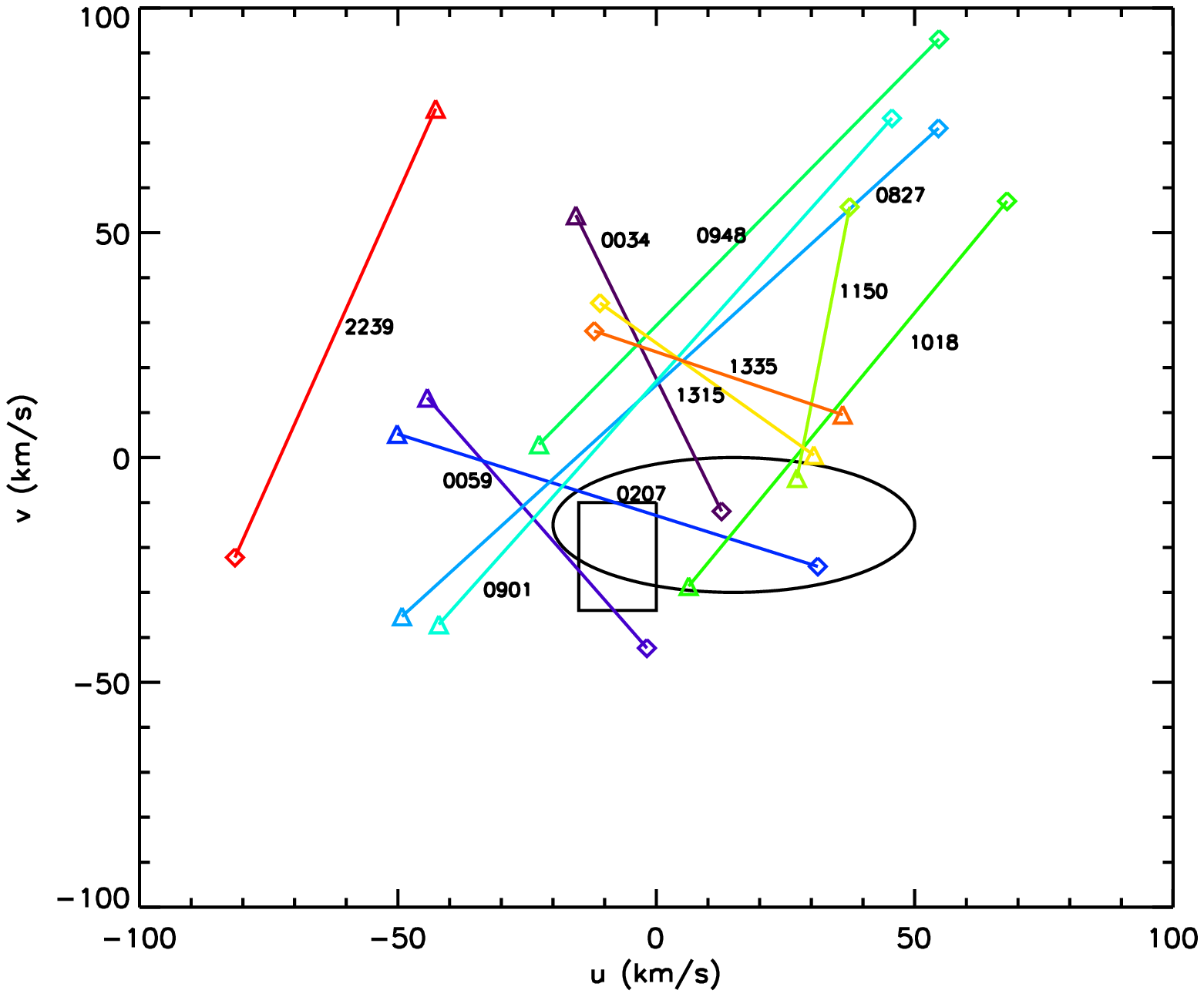}
  \includegraphics[width=8.5cm]{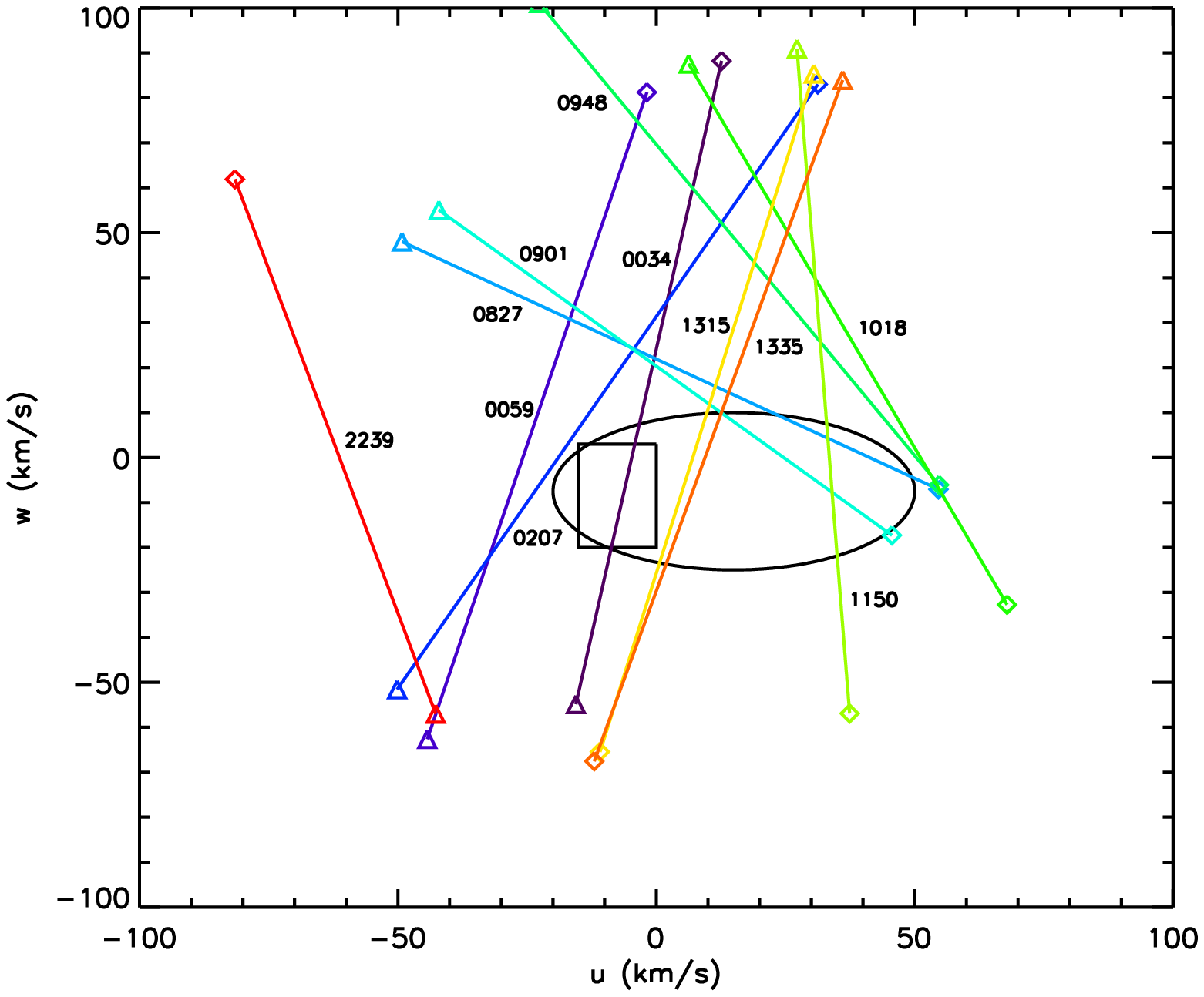}
  \includegraphics[width=8.5cm]{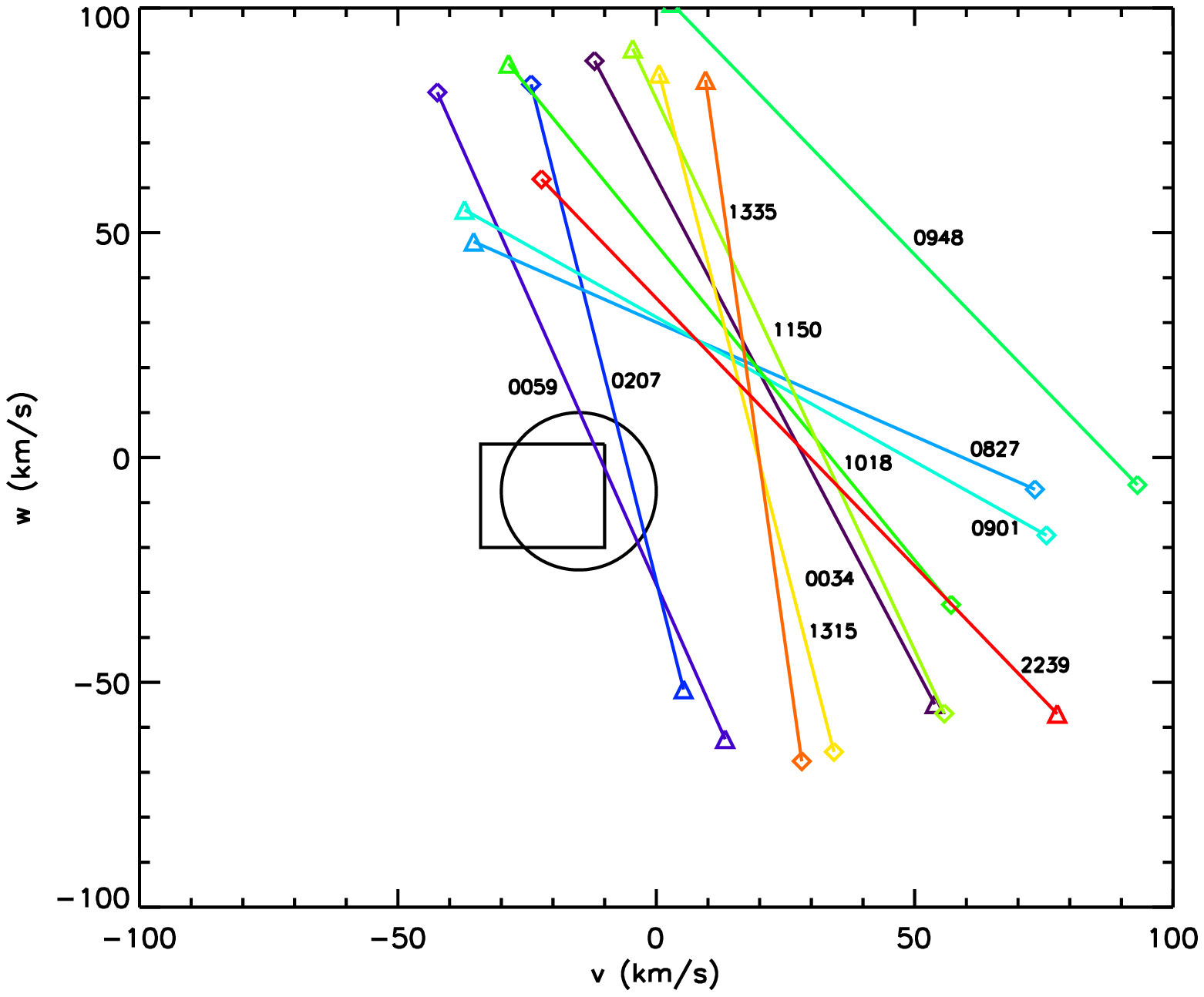}
  \caption{The galactic velocity components U, V and W obtained from the
    proper motions in Table \ref{results} assuming a V$_{rad}$ range of +80/-80 km/s. 
    The triangle indicates the +80 km/s extreme, the diamond indicates the -80 km/s extreme. 
    The overplotted black box is the locus of young stars \citep[age$<$ 0.1 Gyr]{2004ARA&A..42..685Z}, the ellipsoid 
    is the locus of young disk stars \citep[age$<$ 0.5 Gyr]{1969PASP...81..553E}.} 
  \label{uvw}
\end{figure}

To determine a radius, we consider the models of \citet{2003A&A...402..701B}
we see that the radius of T dwarfs decreases rapidly when the object is very
young but after 0.5 Gyr it is very constant. Therefore if we can constrain the
age of our objects we can use these models to constrain the radius. In
Fig. \ref{uvw} we report the galactic velocity components (U,V,W) of the 11
targets. The components were calculated from the proper motions in Table
\ref{results} assuming a radial velocity range of +80/-80 km/s. Given this
large range in V$_{rad}$, the uncertainty in the proper motion becomes
negligible, so we ignore it in this calculation. The overplotted box is the
locus of very young objects \citep[age $<$ 0.1 Gyr,][]{2004ARA&A..42..685Z}
while the ellipsoid is the locus of young disk objects \citep[age $<$ 0.5
Gyr,][]{1969PASP...81..553E}. Even given this very conservative range in
possible radial velocities no objects fall in the boxed area or in the
ellipsoid area in all three components. We therefore conclude that all objects
are older than 0.5 Gyr. For the 0.5-10 Gyr models from
\citet{2003A&A...402..701B} we find the radius has a range of 1.2 - 0.8
R$_{Jup}$ (in a temperature range of 500 - 2000 K). This is also consistent 
with the radii predicted by \citet{1997ApJ...491..856B} models for objects 
of the same age and temperature.

To calculate the bolometric flux (and hence the luminosity) we combined the
available measured spectrum of each object
\citep{2008MNRAS.391..320B,2009ApJ...695.1517L} flux calibrated using UKIDSS
\textit{YJHK} photometry, with the model spectra. For ULAS~0034 and 1335 we have
the spectrum in the near- and mid-infrared (hereafter MIR) region, while for
the other objects we only have the NIR portion. To find the bolometric flux we
use model spectra to estimate the flux where we do not have observations: at
short wavelengths ($\lambda$ $<$ 1.0 $\mu$m), and, for ULAS~0034 and
1335, in the region between the near- and mid-infrared spectrum (2.4
$\mu$m $<$ $\lambda$ $<$ 7.5 $\mu$m) or, for the other targets, in the entire
portion from 2.4 $\mu$m to 15 $\mu$m. The flux emitted beyond 15 $\mu$m was 
estimated assuming a Rayleigh-Jeans tail.

The models used are the already mentioned BTSettl09 and the non-equilibrium
models by \citet{2007ApJ...669.1248H}, covering the temperature range from 700
to 1900 K, for log[g]=4.5,5.0 and 5.5, assuming values of
K$_{zz}$=10$^2$,10$^4$ and 10$^6$ cm$^2$s$^{-1}$ (eddy diffusion
coefficient), for different speed of the CO/CH$_4$ reaction (for further
details see \citealt{2007ApJ...669.1248H,2006ApJ...647..552S} and reference therein).

For each object we took the models in a wide range of temperature ($\pm$200 K
around the temperature predicted by the temperature-type relation given by
\citealt{2009ApJ...702..154S}\footnote{Except for the T9s, since we don't have 
theoretical spectra for temperatures lower than 500 K.}), gravity (log[g]
from 4.5 to 5.5) and metallicity ([Fe/H] from -0.2 to +0.2) and we scaled them
using the available magnitudes (listed in Tables \ref{photometry} and
\ref{irac}). For each model we took the average scaling factor obtained and we compared it with the
range given by the known distance and the radius range adopted (i.e. the square 
of 0.8 R$_{Jup}$/distance $\pm$ 3$\sigma$ - the square of 1.2 R$_{Jup}$/distance 
$\pm$ 3$\sigma$). We discarded the models whose average scaling factor was out 
of this pseudo-3$\sigma$ range. We joined each one of the remaining spectra with
the measured one and we calculate the resulting bolometric flux, the
luminosity and hence the temperature range (corresponding to the radius
range). Then we compared the temperature obtained with the one associated with
the model employed. We discarded those models whose temperature differed by
more than 100 K from the temperature range obtained. Finally we assumed the
mean flux given by the remaining models as our final estimation and hence we
calculate the luminosity and the temperature range.

The uncertainty in the flux is given by the spread in values plus a 3\%
uncertainty in the magnitudes used to calibrate the observed spectra and to
scale the model ones. The uncertainty in the luminosity and the temperature is
the result of the standard propagation of the errors on the flux and the
distance, ignoring the uncertainty in the radius, given the wide range
adopted.

\begin{table}
\centering
\caption{IRAC magnitudes of ULAS~0034, SDSS~0207 and ULAS~1335
  \citep{2007MNRAS.381.1400W, 2006ApJ...651..502P, 2008MNRAS.391..320B} used
  to scale the model spectra.}

\begin{tabular}{c|c|c|c}
   { } & ULAS~0034 & SDSS~0207 & ULAS~1335 \\
\hline
	 3.55 $\mu$m & 16.28 $\pm$ 0.49 & 15.59 $\pm$ 0.06 & 15.96 $\pm$ 0.48 \\
	 4.49 $\mu$m & 14.49 $\pm$ 0.43 & 14.98 $\pm$ 0.05 & 13.91 $\pm$ 0.42 \\
	 5.73 $\mu$m & 14.82 $\pm$ 0.44 & 14.67 $\pm$ 0.20 & 14.34 $\pm$ 0.43 \\
	 7.87 $\mu$m & 13.91 $\pm$ 0.42 & 14.17 $\pm$ 0.19 & 13.37 $\pm$ 0.41 \\
\hline	
\end{tabular}
\label{irac}
\end{table}

The results are shown in Table \ref{temperature}. In the first column we
indicate the target short name, in the second one its spectral type, in the
third the temperature estimated using the temperature-spectral type relation
given by \citet{2009ApJ...702..154S}, in the fourth the range of models
employed, in the fifth the range of models kept after the first step (so after
comparing the scaling factors), in the sixth the range of models kept after
the second step (so after comparing the temperatures), in the seventh our
assumed bolometric flux, in the eighth the associated luminosity and in the
last the temperature range obtained.

We note that the use of MIR magnitudes and spectra increases our ability to constrain the object temperature. For ULAS~0034 and 1335 the bolometric flux is well constrained and the width of the temperature range obtained (150-200 K) is mainly due to the radius. Using the NIR spectrum and photometry only, the uncertainty on the flux increases and the temperature range obtained doubles.

\begin{table*}
\caption{Fluxes, luminosities and temperatures of the sample obtained scaling
  the model spectra using the measured magnitudes, e.g. the first method
  discussed in Section 5.}
\begin{tabular}{c|c|c|c|c|c|c|c|c}
	object & Sp. & est. & models & 1st step & 2nd step & F$_{bol}$ & L/L$_\odot$ & T$_{eff}$ range ($\sigma$)\\
	 & Type & T$_{eff}$ (K) & used (K) & (K) & (K) & (erg s$^{-1}$ cm$^{-2}$) & & (K) \\
\hline
  ULAS~0034 & T9 & 500 & 500:700 & 540:700 & 540:660 & 2.34$\pm$0.05$\times$10$^{-13}$ & 1.13$\pm$0.06$\times$10$^{-6}$ & 535 - 660 (35) \\
  CFBDS~0059 & T9 & 500 & 500:700 & 500:620 & 500:620 & 2.85$\pm$0.48$\times$10$^{-13}$ & 7.20$\pm$1.25$\times$10$^{-7}$ & 480 - 590 (55) \\
  SDSS~0207 & T4.5 & 1130 & 900:1400 & 1000:1400 & 1100:1200 & 4.98$\pm$0.26$\times$10$^{-13}$ & 1.72$\pm$0.25$\times$10$^{-5}$ & 1060 - 1300 (110) \\
  ULAS~0827 & T5.5 & 1070 & 900:1300 & 1000:1300 & 1000:1100 & 2.87$\pm$0.09$\times$10$^{-13}$ & 1.21$\pm$0.15$\times$10$^{-5}$ & 970 - 1190 (95) \\
  ULAS~0901 & T7.5 & 830 & 600:1000 & 600:700 & 600:700 & 1.88$\pm$0.15$\times$10$^{-14}$ & 1.42$\pm$0.13$\times$10$^{-6}$ & 570 - 700 (50) \\
  ULAS~0948 & T7 & 910 & 700:1100 & 700:800 & 700:800 & 6.52$\pm$0.44$\times$10$^{-14}$ & 2.61$\pm$0.44$\times$10$^{-6}$ & 660 - 810 (80) \\
  ULAS~1018 & T5 & 1100 & 900:1300 & 900:1100 & 900:1000 & 1.80$\pm$0.43$\times$10$^{-13}$ & 8.54$\pm$0.70$\times$10$^{-6}$ & 890 - 1090 (75) \\
  ULAS~1150 & T6.5p & 980 & 800:1200 & 800:1200 & 900:1000 & 6.76$\pm$0.40$\times$10$^{-14}$ & 7.12$\pm$3.22$\times$10$^{-6}$ & 850 - 1040 (160) \\
  ULAS~1315 & T7.5 & 830 & 600:1000 & 600:740 & 600:660 & 7.47$\pm$0.72$\times$10$^{-14}$ & 1.21$\pm$0.25$\times$10$^{-6}$ & 545 - 670 (70) \\
	ULAS~1335 & T9 & 500 & 500:700 & 540:660 & 540:660 & 3.42$\pm$0.09$\times$10$^{-13}$ & 1.09$\pm$0.06$\times$10$^{-6}$ & 530 - 650 (35) \\
  ULAS~2239 & T5.5 & 1070 & 900:1300 & 900:1300 & 1100:1200 & 6.00$\pm$0.09$\times$10$^{-14}$ & 1.65$\pm$0.82$\times$10$^{-5}$ & 1050 - 1280 (200) \\
\hline
\end{tabular}
 \\
\label{temperature}
\end{table*}

To have an idea of the eventual systematics, we also tested the technique described in \citet{2008ApJ...678.1372C}, e.g. to fit the object spectrum using the model spectra. The best fit spectrum was selected as the one that minimize:
\begin{equation}
 G_k = \sum_{i=1}^n w_i \left( \frac{f_i-C_k{}F_{k,i}}{\sigma_i} \right)^2
\label{c1}
\end{equation}
where \emph{n} is the number of data pixels, f$_{i}$ is the measured flux in
the i-th spectral interval, C$_k$ is the scaling factor (R/d)$^2$, F$_{k,i}$
is the k-th model flux and $\sigma${}$_i$ is the error in the measured
spectrum. The weight associated to each bin (w$_i$) is the extension of the
bin itself ($\Delta${}$\lambda$) as suggested by Cushing et al. The scaling
factor can be provided by the fit, however, since we know the distance to the
dwarf we consider fixed values of C$_k$, assuming again the radius range
previously indicated. We selected the best fit models for the two extreme
configurations, i.e. 0.5 Gyr-1.2 R$_{Jup}$ and 10 Gyr-0.8 R$_{Jup}$.

Since we don't have an associated noise spectrum for ULAS~0827, 0948, 1018, 1150, 1315 and CFBDS~0059 we minimize:
\begin{equation}
 G_k = \sum_{i=1}^n w_i \left( \frac{f_i-C_k{}F_{k,i}}{\sqrt{f_i}} \right)^2
\label{c2}
\end{equation}

The uncertainty in the extremes takes into account half of the models grid
spacing based on fitting (for further details see Cushing et al.), plus an
additional percentage on the flux due to the incomplete spectral
coverage. This percentage was estimated comparing the measured IRAC magnitudes
of ULAS~0034 and 1335 with the model's predicted ones. The differences between
measured and model magnitudes gives an average uncertainty of $\sim$70$\%$ on
the calculated flux between 2.5 and 7.5 $\mu$m. For ULAS~0034 and 1335 in this
interval there is $\sim$40$\%$ of the total emergent flux, so we obtain a
relative error $\sigma${}$_{F,rel}$=.7$\times$.4=.28. This implies an
additional uncertainty of $\sim$7$\%$ on the temperature. For the other 9
objects, the uncertainty in the temperature was estimated extending the
relative sigma calculated for ULAS~0034 and 1335 to the uncovered part of the
flux ($\sim$60$\%$, that results in an uncertainty
$\sigma${}$_{F,rel}$=.7$\times$.6=.42).

The choice of the weight function is arbitrary. Different choices can lead to different results, as seen by \citet{2008ApJ...678.1372C} and \citet{2009ApJ...702..154S}. Given this, we prefer the results obtained with the first method described. The values obtained with this spectral technique are summarized in Table \ref{teff_fit}. They are largely consistent with the ones obtained scaling the model spectra. 

We also performed a completely model independent flux calculation for ULAS~0034 and 1335, that have a completely measured spectral energy distribution. We determined the bolometric flux emitted integrating the measured spectrum between 1 and 2.5 $\mu$m, then adding the flux emitted between 2.5 and 7.5 $\mu$m calculated using the IRAC magnitudes (assuming a constant flux distribution over the passband\footnote{Any error that this assumption may introduce would be negligible, since the IRAC passbands are tight.}), finally we integrated the measured 7.5 - 15 $\mu$m spectrum. The flux emitted beyond 15 $\mu$m was estimated assuming a Rayleigh-Jeans tail. The results obtained for ULAS~0034 and 1335 with this approach are consistent with the one obtained with the other techniques described within the uncertainty quoted in Table \ref{temperature}.

\begin{table}
\centering
\caption{Temperatures obtained fitting the observed spectrum, e.g. the second method
  discussed in Section 5.}
\begin{tabular}{c|c|c}
Name & Sp. Type & T$_{eff}$ range ($\sigma$)\\
 & & (K) \\
\hline
ULAS 0034  & T9    & 500 - 580 (40) \\
CFBDS 0059 & T9    & 500 - 540 (60) \\
SDSS 0207  & T4.5  & 1000 - 1200 (120) \\
ULAS 0827  & T5.5  & 1100 - 1200 (130) \\
ULAS 0901  & T7.5  & 620 - 720 (75) \\
ULAS 0948  & T7    & 700 - 800 (90) \\
ULAS 1018  & T5    & 900 - 1100 (120) \\
ULAS 1150  & T6.5p & 800 - 1000 (110) \\
ULAS 1315  & T7.5  & 540 - 620 (65) \\
ULAS 1335  & T9    & 500 - 600 (40) \\
ULAS 2239  & T5.5  & 1100 - 1300 (135) \\
\hline
\end{tabular}
 \\
\label{teff_fit}
\end{table}

\section{Discussion}
\begin{figure}
  \centering
  \includegraphics[width=8.5cm]{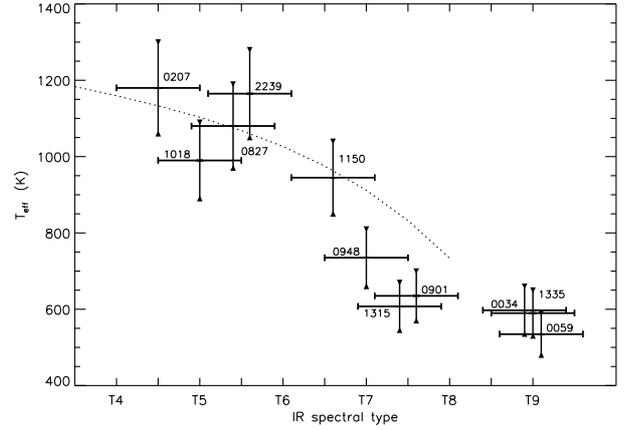}
  \caption{Temperature ranges plotted as a function of the spectral type. Uncertainties in each extreme point of the range are in Table 
  \ref{temperature}. The uncertainty in spectral type is half subtype. The over plotted dotted line is the effective temperature-infrared type relation derived by \citet{2009ApJ...702..154S}.} 
  \label{teff-vs-sptype}
\end{figure}

In Fig. \ref{teff-vs-sptype} we present a T$_{eff}$-spectral type diagram of 
the 11 T-dwarfs of the sample, for each of which we plot the temperature range 
displayed in Table \ref{temperature}. When needed, objects have been offset 
by $\pm$0.1 in spectral type, to avoid overlaps. Over plotted for comparison 
we have the effective temperature - infrared type relation derived by 
\citet{2009ApJ...702..154S}. All the ranges are consistent with the relation 
except for ULAS~0948, 0901 and 1315, that are cooler than predicted.

We now discuss individual objects with temperature values or indications of peculiarity in the literature.

\subsection{ULAS~0034} 
There are several estimations of the temperature of this object. In the
discovery paper, \citet{2007MNRAS.381.1400W} estimate a conservative range of
600 $\leq$ T$_{eff}$ $\leq$ 700 K using the measured  NIR
spectrum and the near- and mid-infrared photometry with a grid of
solar-abundance BTSettl model, calibrated using the parameters of 2MASS0415,
and a linear fit of T$_{eff}$ vs. H-[4.5] color for hotter stars;
\citet{2008AA...482..961D} using the NIR spectrum and the same grid described
in \citet{2007MNRAS.381.1400W} determine T$_{eff}$ $\approx$ 670 K; in
\citet{2009ApJ...695.1517L} the published range is 550 $\leq$ T$_{eff}$ $\leq$
600, obtained fitting the measured NIR and MIR spectrum with model spectra by
\citet{2008ApJ...689.1327S}; finally, in SJL10 using the parallax reported in 
this paper and the model fitting in \citet{2009ApJ...695.1517L}, using a 
model radius of 0.11 R$_\odot$ (as implied by the spectral fits
constraints on gravity), derive a bolometric luminosity of 
L/L$_\odot$=1.10$\pm$0.01$\times$10$^{-6}$, consistent with the temperature 
range 550-600 K. In this work, using NIR and MIR spectrum, the MKO NIR 
photometry and the IRAC MIR photometry, we find a luminosity of 
L/L$_\odot$=1.13$\pm$0.06$\times$10$^{-6}$, consistent with the value obtained 
by \citet{2009ApJ...695.1517L}. The difference between the temperature range 
derived here and by Leggett et al. is due to the different technique used: 
the Leggett et al. spectral fitting result is similar to that derived 
spectrally here, given in Table \ref{teff_fit}.

\subsection{CFBDS~0059}
In the discovery paper \citep{2008AA...482..961D} of this object they estimate
T$_{eff}$$\sim$620~K comparing the spectral indices with a solar metallicity
grid of BTSettl model spectra. In \citet{2009ApJ...695.1517L} the technique
adopted is the same described in Sec. 6.1, and the range obtained is
550$<$T$_{eff}$$<$600. Here we found a range 490$<$T$_{eff}$$<$600 consistent
with Leggett et al., larger because of the different technique used. Moreover, 
as noted in \citet{2008AA...482..961D} using the Besan\c con stellar population 
model \citep{2003A&A...409..523R}, the kinematics suggest that the target is a 
older member of the thin disk (age$\sim$4 Gyr) that corresponds, using 
\citet{2003A&A...402..701B} radii, to a radius R$\approx$0.9 R$_{Jup}$ and a 
T$_{eff}$$\approx$570 K. As regards its gravity and metallicity, we noted in 
Sec. 4 that the model predictions are not consistent, so we do not go further 
in the analysis of these physical properties.

\subsection{SDSS~0207}
In \citet{2004AJ....127.2948V} they find a $\pi_{abs}$=34.85$\pm$9.87 mas and
$\mu_{tot}$=156.3$\pm$11.4 mas/yr for SDSS~0207, both consistent with the values found here.
The model predictions in Figs. \ref{mvsj-h+model} and \ref{mvsj-k+model}
indicates log[g]=5.0 and low or solar metallicity while the BSH06 clear tracks
indicates a metal-rich nature in J-K.

There are two previous estimates of effective temperature for this object:
\citet{2004AJ....127.3516G}, using the measured spectrum and estimating the
bolometric correction using M' and L' photometry, find
L$_{bol}$/L$_\odot$=1.51$\pm$0.87$\times$10$^{-5}$;
\citet{2004AJ....127.2948V} using the K-band magnitude and the bolometric
correction derived by Golimowski et al. and find
L$_{bol}$/L$_\odot$=1.74$\pm$1.04$\times$10$^{-5}$. Both these values are
consistent with our result.  

\subsection{ULAS~0901}
ULAS~0901 is much cooler than what expected for its type, see Fig. 
\ref{teff-vs-sptype}. In Fig. \ref{magvsirsptype} the dwarf is fainter 
than the other T7.5s of the sample, in particular in M$_J$ and M$_H$. 
This maybe an indication that ULAS~0901 is half a subclass later, i.e. a T8.

In \citet{2007MNRAS.379.1423L} ULAS~0901 is indicated as a high gravity - 
solar metallicity object, based on the low K-band flux. The model prediction 
of the gravity and the metallicity of this object in Figs. \ref{mvsj-h+model} 
and \ref{mvsj-k+model} are log[g]=4.5 and [Fe/H]=+0.2 (in contrast with this 
finding) but we must point out the known degeneracy between gravity and 
metallicity \citep{2004AJ....127.3553K}. This degeneracy could be the reason 
for the discrepancy between our result and Lodieu et al.

\subsection{ULAS~1018}
The model predictions in Figs. \ref{mvsj-h+model} and \ref{mvsj-k+model} are 
not consistent for ULAS~1018. In the J-K space it appears as a solar-metallicity low-gravity 
(log[g]=4.5) object, while the J-H prediction is low-gravity and low-metallicity 
([Fe/H]=-0.5). The uncertainty in color and magnitude prevent us to draw 
a firm conclusion on the properties of this object. We mention here that ULAS~1018 
is indicated as a metal-poor object by Murray et al. (in prep.) based on its very 
blue H-K, while in \citet{2007MNRAS.379.1423L}, was indicated as a possible metal-rich 
object, based on the high K-band flux and narrow Y-band flux peak.

The bolometric luminosity and thus the temperature range obtained are consistent with an 
object of spectral type T5.

\subsection{ULAS~1150}
This object is a peculiar T6.5, with T7-T6.5 indices in the J and K band
\citep[H$_2$O-J = 0.087, CH$_4$-J = 0.302, CH$_4$-K =
0.032,][]{2008MNRAS.390..304P}, but T3 in the H (H$_2$O-H = 0.455). Pinfield et
al. suggest that it could be a low-gravity high-metallicity object, based
on the K-band enhancement and the Y-band suppression. In Fig. 
\ref{mvsj-h+model} it assumes an anomalous position, lying on the low-gravity 
high-metallicity side of the theoretical curves  while in Fig.\ref{mvsj-k+model} 
the model predicts a low/solar metallicity. The large errors prevent us from
making any comments on its gravity.

The bolometric luminosity and thus the temperature range obtained are consistent with an 
object of spectral type T6.5.

\subsection{ULAS~1315}
In Fig. \ref{teff-vs-sptype} it appears significantly cooler than the other
T7.5s. In the discovery paper, \citet{2008MNRAS.390..304P}, no
particular indication of peculiarity were detected. Looking at
Figs. \ref{mvsj-h+model} and \ref{mvsj-k+model} it is easy to see that
ULAS~1315 is particularly faint in M$_K$ (it is the bluest in J-K), and the
K-band suppression indicates high gravity, as predicted by the models. However
the indication of high metallicity in the same plots is in contradiction with
this interpretation as a metal-rich object would show an enhancement in the
K-band flux.
It is fainter than the other T7.5s also in M$_J$ and M$_H$, this results in an 
extremely cool temperature, almost 200 K less than what expected. This can indicate 
that ULAS~1315 is a T8 (the spectral indices H$_2$O-J, CH$_4$-J and CH$_4$-H corresponds 
to a T8, while H$_2$O-H to a T7, see \citealt{2008MNRAS.390..304P}).

\subsection{ULAS~1335}
There are two previous estimations of the temperature of this object:
\citet{2008MNRAS.391..320B}, using a solar-metallicity BTSettl model grid and
a comparison of the H-[4.49] color with theoretical expectations, find
T$_{eff}$$\sim$550-600 K; \citet{2009ApJ...695.1517L}, using the technique
described in Sec. 6.1, find T$_{eff}$=500-550 K. The luminosity obtained here
(L$_{bol}$=1.09$\pm$0.06$\times$10$^{-6}$ L$_\odot$) is formally consistent
with both those ranges, given the large uncertainty in the radius, but it is
closer to the value obtained in \citet{2008MNRAS.391..320B}. As in the case of
the other late T dwarfs, ULAS~0034 and CFBDS~0059, the model predictions are
contradictory in the different color spaces and we are therefore unable to say
anything about the other parameters of this objects.

\section{Conclusions}
We present the parallax and proper motions of 11 cool T dwarfs taking
advantage of the UKIDSS discovery image to shorten the time required for a
precise determination. We find that the models do not predict the colors and
absolute magnitudes of the coolest T dwarfs, and that, given the observed 
colors and absolute magnitudes, the models predict contradictory behaviours 
depending on which color is considered.

We examine two methods for the calculation of T$_{eff}$, one of which derives 
the temperature from the luminosity, which is determined using available spectroscopic 
data complemented with model spectra scaled using measured photometry, and the other 
of which does a least-squares-like model spectrum fit to the observed spectrum. The 
second method is very dependent on the data weighting selection and so we prefer the 
former technique. The two approaches give consistent results however the 
former gives a larger range in temperature.

The observations of these 11 objects are continuing at a lower
frequency of 4 observations per year. At the end of the observing campaign
all the objects in the sample will have a time coverage of 4 years, allowing
us to obtain a more precise and robust parallax solution.  Also, enhanced
photometry for ULAS~0901 and ULAS~0948 is expected within the year.

The inclusion of MIR observations increases our ability to determine the object parameters, 
especially for the temperature. With IRAC photometry and a MIR spectrum the emitted flux is 
well constrained and the limiting factor becomes the radius, producing a range of $\sim$150 K. 
Using a NIR spectrum and photometry only, the uncertainty in the flux is on the order of that 
on the radius, and the temperature range obtained doubles. For the future we hope that the 
Warm-Spitzer and WISE missions will continue to provide MIR observations for the study of 
these objects.

\begin{acknowledgements}
  The authors would like to acknowledge the support of: the Royal Society
  International Joint Project 2007/R3; the PARSEC International Incoming
  Fellowship and IPERCOOL International Research Staff Exchange Scheme within
  the Marie Curie 7th European Community Framework Programme. This research
  has benefitted from the M, L, and T dwarf compendium housed at
  dwarfArchives.org and maintained by Chris Gelino, Davy Kirkpatrick, and Adam
  Burgasser.  The United Kingdom Infrared Telescope is operated by the Joint
  Astronomy Centre on behalf of the Science and Technology Facilities Council
  of the U.K., all of the data used here were obtained as part of the UKIRT
  Service Programme.
  SKL's research is supported by the Gemini Observatory, which is operated by 
  the Association of Universities for Research in Astronomy, Inc., on behalf 
  of the international Gemini partnership of Argentina, Australia, Brazil, 
  Canada, Chile, the United Kingdom, and the United States of America.
  NL acknowledges funding from the Spanish Ministry of Science and
  Innovation through the Ram\'on y Cajal fellowship number 08-303-01-02\@.
\end{acknowledgements}

\bibliographystyle{aa} 
\bibliography{15394}

\end{document}